\documentclass[prb,twocolumn,groupedaddress,letterpaper,showpacs]{revtex4}

\usepackage{graphicx}

\begin{document}

\title{Dynamics of entanglement in two coupled qubits}

\author{Vivek M. Aji}
\author{J. E. Moore}

\affiliation{Department of Physics, University of California at
Berkeley, Berkeley, CA 94720}

\affiliation{Materials Sciences Division, Lawrence Berkeley National Laboratory, Berkeley, CA 94720}

\date{\today}

\begin{abstract}
Considerable progress has been made in understanding how dissipation drives the state
of a single qubit from a quantum-mechanical superposition to a classical mixed state.  This paper uses a
Bloch-Redfield approach to study how, in a system of two qubits, dissipation
drives the bipartite state of the two qubits from an entangled state to a product state.  The measure used
for mixed-state entanglement is Wootters's formula for the entanglement of formation.
For qubits that start in an entangled state and are then decoupled, entanglement is found to decay faster in general than the decoherence of single qubit states by a factor $2/\log 2$ that can be obtained analytically in a limiting case.  We show that the dynamics of entanglement for realistic parameters is different from that of fidelity: for some experimental parameters a bipartite state loses fidelity much more rapidly than it loses entanglement.  These results are of some practical interest since entanglement is thought to be the essential ``fuel'' for fast quantum algorithms. 
\end{abstract}

\pacs{85.25.-j, 85.25.Cp, 85.25.Dq}

\maketitle

\section{Introduction}
Over the past few years considerable progress has been made in fabricating and manipulating solid state
qubits \cite{Chio, Yu, Nakamura, Vion, YYu, Mart, Fried}. A particular implementation of interest involves superconducting elements where either the charge or flux states are exploited for the purposes of quantum computation \cite{Aver, Makh, Orl, Moo}. Other systems that are also being actively studied include trapped ions \cite{Cir, Mon, Poy, Leib, Kale}, and nuclear moments of atoms in an NMR setup \cite{Div, Chu, Linden}. One of the requirements for the purposes of quantum computation is the ability to isolate qubits from external noise. Unfortunately perfect isolation is never possible in practice, as the solid state qubits are always coupled to their environment, but much experience has been gained in observing quantum phenomena in macroscopic solid state systems\cite{Clarke1, Clarke2}. The processes of preparing, manipulating and measuring the state of the qubits also provide mechanisms for decoherence and dissipation.

The standard theoretical approach to decoherence in a single qubit models the environment as a bath of oscillators coupled to the qubit\cite{Cald}: the dissipative environment causes a coherent superposition in the qubit to evolve toward a mixed state. Current interest in the field of quantum computation is focused on fabricating coupled qubit systems so as to be able to perform nontrivial computations based on quantum algorithms. Of great importance in multi-qubit systems is the notion of entanglement, a characteristic trait of quantum mechanics, which is believed to be the source of the exponential speed-up of quantum algorithms\cite{jozsa}.  In this paper we study the time evolution of entangled states in the presence of a dissipative environment: a fundamental question about coupled qubit systems is how dissipation causes entangled states to evolve into product states.

For single qubits, the quantum nature of the system is investigated by looking at either the Rabi oscillations or Ramsey fringe measurement, both of which rely on the quantum coherence of the state in question. The presence of an environment leads to loss of quantum coherence, or dephasing, which limits the operation of the qubit as a device for computation. On the other hand, for coupled qubits, the key quantum signature is entanglement and our primary motivation is to investigate the two qubit system with entanglement as the diagnostic. We find interesting temporal behavior of the entanglement of formation~\cite{Woot} in coupled systems, and find that once coupling is turned off, entanglement decays at a rate that is roughly three times faster than the single qubit decoherence rate.  This may prove relevant for implementing logic gates in quantum computers.

Entanglement is a measure of the amount of information of one qubit that can be obtained by making a measurement on the second of a pair of coupled qubits.  To illustrate consider the singlet state of a two spin system: $\left| \psi \right> = (1/\sqrt{2})(\left| \uparrow \downarrow \right> - \left| \downarrow \uparrow \right>)$. Measurement of the state of one, completely determines the outcome of a measurement performed on the other. Thus, the notion of entanglement. Such a determination would not be possible if the state of the two spin system could be represented as a product state of the two individual spins, i.e. $\left| \psi_{1}\right>\left|\psi_{2}\right>$: any product state has zero entanglement.  In particular the (maximally entangled) singlet state cannot be represented as a product of individual states. The two states of the spins can be identified with the $0$ and $1$ states of a qubit.

For a bipartite system in a pure quantum state, the von Neumann entropy of either of the pair can be used as a measure of entanglement, $E(\psi) =-$Tr($\rho$ Log$_{2}$ $ \rho$), where $\rho$ is the partial trace of $\left| \psi \right>\left< \psi \right|$ over either of the subsystems. For a mixed state the quantification is more complicated, because the definition of entanglement must distinguish between quantum entanglement and classical correlation.  While the singlet state can still be used as the basic unit of entanglement, the number of singlets needed to create a mixed state is different from the number one can extract from the state.  (By extracting we mean the number of singlets produced by local operations and classical communication from an entangled state $\psi$.) For the purposes of quantitative calculation of entanglement dynamics, we will follow the prescription given by Wootters for the entanglement of formation \cite{Woot},
reviewed in Section III below.

Gate performances and coherent dynamics in coupled qubit systems have previously been studied\cite{Stor, Gov}, where the coupling between the two is of the form $\sigma_{z}^{(1)} \otimes \sigma_{z}^{(2)}$ (Ising coupling) and $\sigma_{y}^{(1)} \otimes \sigma_{y}^{(2)}$ respectively, $\sigma^i_j$ being the Pauli matrices describing the two-state quantum system $i$.  (In this notation it is $\sigma_z$ that couples to the dissipative bath.)  The models we introduce in Section II are generally similar to these, although we find that the dynamics to be more interesting in the case where the coupling has $SU(2)$ symmetry, i.e., is of the form $\vec{\sigma}^{(1)} \otimes \vec{\sigma}^{(2)}$, as is likely appropriate for spin qubits.  However, the primary emphasis of our work is the study of entanglement rather than fidelity of quantum evolution processes of two qubits.  We find that some operations destroy fidelity but preserve entanglement, while others destroy both fidelity and entanglement; hence the two are sensitive to different physical processes.

Preservation of entanglement is essential for applications because lost entanglement cannot be restored by local operations and classical communcation.  Sections IV and V study how entanglement evolves in the models of Section II, for a variety of initial states and couplings.  In the limit of zero coupling the entanglement is seen to decay due to the single qubit decoherence, while at finite coupling the system can show nonmonotonic behavior where the initial entanglement is lost at intermediate times, but is regained as the system evolves further. This behavior is indicative of a transition from one entangled state to another caused by dissipation, which can be inferred by the time evolution of the fidelity.

\section{The Model}

Each individual qubit is a two state system that we describe in the pseudospin representation along the $z$ axis: the two states are $|\uparrow\rangle, |\downarrow\rangle$.  For flux qubits, the bias $e$ along this direction is proportional to the net flux, $\Phi$, through the superconducting loop making up the qubit, $ e = 2I_{p}(\Phi - {1\over 2}\Phi_{0}) $, where $I_{p}$ is the persistent current in the loop and $\Phi_{0}$ is the flux quantum. In addition there exists a tunnel splitting, $\Delta$, between the two states with currents flowing in opposite directions at the degeneracy point, $\Phi = \Phi_{0}/2$. This is captured in the Hamiltonian with a term proportional to the $x$-component of the pseudospin. Similar representations for the Hamiltonians for the NMR and ion trap realizations for qubits can be developed where the bias is the $z$ component of the applied static magnetic field and the tunneling $\Delta$ is the time dependent transverse magnetic field, which has components both in the $x$ and $y$ directions. 

The environment is modeled by a bath of oscillators which is coupled to the individual qubits. The form of the coupling depends on the system in question. For flux qubits, the noise in the flux threading the superconducting loop is the primary source of dissipation. Hence the Hamiltonain includes a term that couples the oscillator degrees of freedom to that of the $z$-component of the pseudospin. Similar models would apply for the NMR and ion trap systems if noise in the static applied field is the primary source of decoherence. The final piece of the model is the coupling between the two qubits. For NMR setups, there are two forms of interaction; the spin dipolar coupling and the through-bond interaction (or $J$-coupling). The former averages away leaving only a Heisenberg-type interaction between the two spins. For flux qubits inductive coupling leads to term of the type $\sigma_{y}\otimes\sigma_{y}$ \cite{Yu1}, while for other superconducting realizations a coupling of the form $\sigma_{z}\otimes\sigma_{z}$ is appropriate. Here we study the Heisenberg and Ising cases as representative examples.  Putting all the pieces together, we arrive at the following Hamiltonian (with Heisenberg coupling)

\begin{eqnarray}
H &=& H_{q} + H_{bath}+H_{qq} + H_{q-bath}\\ \nonumber
H_{q} &=& -{1 \over 2}e_{1} \hat{\sigma}_{z}^{(1)} - {1 \over 2}\Delta_{1} \hat{\sigma}_{x} ^{(1)} -{1 \over 2}e_{2} \hat{\sigma}_{z}^{(2)} - {1 \over 2}\Delta_{2} \hat{\sigma}_{x} ^{(2)} \\ \nonumber
H_{bath} &=& \sum_{i}\left( {1\over {2m_{i}}}\hat{P_{i}^{2}} + {1\over 2}m_{i}\omega_{i}^{2}\hat{X_{i}^{2}}\right)\\ \nonumber
H_{qq} &=& {J\over 2}\hat{\sigma}^{(1)}\cdot\hat{\sigma}^{(2)}\\ \nonumber
H_{q-bath} &=& {1\over 2}\sum_{i}c_{i}\hat{\sigma_{z}}^{(1)}\hat{X_{i}}+{1\over 2}\sum_{i}c_{i}\hat{\sigma_{z}}^{(2)}\hat{X_{i}}
\end{eqnarray}
where $e_{i}$'s are the biases, $\Delta_{i}$'s are the tunnel splittings, $J$ is the coupling strength and $m$ and $\omega$
characterize the bath of harmonic oscillators coupled to the two qubits. Individual qubits are biased such that $e_{i} \approx 5 \Delta_{i}$. We will further simplify to the case where the two qubits are identical. The eigenvalues are 
(${1\over 8}J-{1\over 2}\sqrt{e^{2}+2n^{2}}, -{3\over 8}J,{1\over 8}J,{1\over 8}J+{1\over 2}\sqrt{e^{2}+2n^{2}}$), where $e = e_{1}+ e_{2}$ and $n =(\Delta_{1}+ \Delta_{2})/\sqrt{2}$. The corresponding eigenvectors in the singlet triplet basis ($\left|\uparrow\uparrow\right>, (\left|\uparrow\downarrow\right> + \left\downarrow\uparrow\right>)/\sqrt{2}, \left|\downarrow\downarrow\right>, (\left|\uparrow\downarrow\right> - \left\downarrow\uparrow\right>)/\sqrt{2}$) are,

\begin{eqnarray}
\left| 1 \right> &=& ({1\over 2}(1+\sin\theta),{1\over{\sqrt{2}}}\cos\theta,{1\over 2}(1-\sin\theta),0)\\ \nonumber
\left| 2 \right> &=& (0,0,0,1)\\ \nonumber
\left| 3 \right> &=& (-{1\over {\sqrt{2}}}\cos\theta, \sin\theta,{1\over{\sqrt{2}}}\cos\theta,0)\\ \nonumber
\left| 4 \right> &=& ({1\over 2}(1-\sin\theta),-{1\over{\sqrt{2}}}\cos\theta,{1\over 2}(1+\sin\theta),0)
\end{eqnarray}
where $\tan\theta = e/\sqrt{2}n$. Note that in the limit $\Delta = 0$, the eigenvectors are the singlet and the triplets themselves. For purposes of our later discussion, we will continue to refer to them as the singlet and triplet, as the operation point for individual qubits is such that $\Delta/e \ll 1$. We will also present results for the case where the interaction is of the form $H_{qq} = K \sigma_{z}^{(1)}\otimes \sigma_{z}^{(2)}$

\section{Entanglement}
To study entanglement, we will follow the prescription of Wootters \cite{Woot}. Given a mixed state $\rho$ of two quantum systems A and B, consider all possible ways of expressing the density matrix, $\rho$, as an ensemble of pure states. For states $\left| \psi_{i}\right>$ and probabilities $p_{i}$, 

\begin{equation}
\rho = \sum_{i}p_{i}\left|\psi_{i}\right>\left<\psi_{i}\right|.
\end{equation}
The entanglement of formation is defined as the minimum, over all such ensembles, of the average entanglement of the pure states making up the ensemble
\begin{equation}
E = \min \sum_{i}p_{i}E(\psi_{i}).
\end{equation}
To avoid the process of extremization, Wootters derived a formula for entanglement \cite{Woot}. Consider the density 
matrix $\rho$ and its complex conjugate $\rho^{*}$ taken in the standard basis for a pair of spin$-{1\over 2}$
systems ($\left|\uparrow \uparrow\right>, \left|\uparrow\downarrow\right>, \left|\downarrow\uparrow\right>, \left|\uparrow\uparrow\right>$). Define the spin flipped state as

\begin{equation}
\widetilde{\rho} = (\sigma_{y}\otimes\sigma_{y})\rho^{*}(\sigma_{y}\otimes\sigma_{y}).
\end{equation}
The entanglement of formation is given by
\begin{eqnarray}\label{ent}
E(C(\rho)) &=& h\left( {{1+\sqrt{1-C^{2}}}\over 2}\right)\\ \nonumber
h(x) &=& -x\log_{2}x - (1-x)\log_{2}(1-x)
\end{eqnarray}
where
\begin{equation}
C(\rho) = \max[0,\lambda_{1}-\lambda_{2}-\lambda_{3}-\lambda_{4}]
\end{equation}
and the $\lambda_{i}$s are the eigenvalues, in decreasing order, of the Hermitian matrix
\begin{equation}
R = \sqrt{\sqrt{\rho}\widetilde{\rho}\sqrt{\rho}}.
\end{equation}

To get a better understanding of this definition of entanglement, consider first pure states. We will work in the following basis (Bell basis),
\begin{eqnarray}
\left| e_{1}\right> &=& {1\over {\sqrt{2}}}\left( \left| \uparrow \uparrow \right> + \left| \downarrow \downarrow \right> \right)\\ \nonumber
\left| e_{2}\right> &=& i {1\over {\sqrt{2}}}\left( \left| \uparrow \uparrow \right> - \left| \downarrow \downarrow \right> \right)\\ \nonumber
\left| e_{3}\right> &=& i {1\over {\sqrt{2}}}\left( \left| \uparrow \downarrow \right> + \left| \downarrow \uparrow \right> \right)\\ \nonumber
\left| e_{4}\right> &=&  {1\over {\sqrt{2}}}\left( \left| \uparrow \downarrow \right> - \left| \downarrow \uparrow \right> \right).\\ \nonumber
\end{eqnarray} 
Consider a pure state $\left| \psi \right> = \sum_{i=1}^{4}\alpha_{i}\left| e_{i}\right>$. The von Neumann entropy of the reduced density matrix of either one of the qubits is given by (\ref{ent}) where the concurrence, $C$, is $C = \left| \sum \alpha_{i}^{2} \right|$ \cite{Ben}. The same result is obtained by using the definition of $C$, via the eigenvalues of the matrix $R$. Furthermore, the trace of the matrix $R$ is a measure of ``distance'' between the states $\rho$ and $\widetilde{\rho}$, which in turn indicates how closely the state $\rho$ approximates a mixture of bell states \cite{Bur, Joz}. Thus the measure of entanglement reviewed here is a generalization of the von Neumann entropy for the pure case.

\section{Time Evolution of the Density Matrix}

In the presence of dissipation the time evolution of the density matrix can be computed using the Bloch-Redfield formalism \cite{Argy}. The equation of motion for the $\rho$ is \cite{Weiss} (Here we follow the notation in \cite{Stor})
\begin{equation}
{\partial\rho_{nm}(t)\over {\partial t}} = -i\omega_{nm}\rho_{nm}(t) - \sum_{kl}R_{nmkl}\rho_{kl}(t)
\end{equation}
where $\omega_{nm} = (\lambda_{n}-\lambda_{m})/\hbar$ and max$_{nmkl}\left| Re(R_{nmkl})\right| < $ min$_{n\neq m}\left|\omega_{nm}\right|$.The golden rule rates yield the following expression for the Redfield tensor
\begin{equation}
R_{nmkl} = \delta_{lm}\sum_{r}\Gamma_{nrrk}^{(+)} + \delta_{nk}\sum_{r}\Gamma_{lrrm}^{(-)} - \Gamma_{nmkl}^{(+)} - \Gamma_{nmkl}^{(-)}.
\end{equation}
With Ohmic dissipation with a Drude cutoff ($\omega_{c}$), the spectral function is
\begin{equation}
J_{i} = {\alpha_{i}\hbar\omega \over {1 + \omega^{2}/\omega_{c}^{2}}}.
\end{equation}
The formalism is valid provided $\alpha_{i} \ll 1$. Notice we have assumed that each qubit is individually coupled to an Ohmic bath. The rates are given by
\begin{eqnarray}
{\Gamma_{lmnk}^{(+)}} &=& {{1 \over {8\hbar}}\left[ \Lambda^{1} J_{1}(\omega_{nk}) + \Lambda^{2}J_{2}(\omega_{nk}) \right] \left[ \coth({{\beta \hbar \omega_{nk}} \over 2}) -1 \right]} \nonumber \\
&+& {{i\over 4\pi\hbar}\left[ \Lambda^{2}M^{(-)}(\omega_{nk}, 2) + \Lambda^{1}M^{(-)}(\omega_{nk}, 1)\right]}
\end{eqnarray}
where $\Lambda^{i} = \sigma_{z,lm}^{(i)} \sigma_{z,nk}^{i}$, $\sigma_{z,lm}^{(i)}$ being the matrix element of $\hat{\sigma}_{z}^{(i)}$, and 
\begin{equation}
M^{\pm}(\Omega, i) = {\cal P}\int_{0}^{\infty} d\omega{J_{i}(\omega) \over {\omega^{2} - \Omega^{2}}}\left[ \coth({\beta\hbar\omega\over 2}) \pm 1 \right]
\end{equation}
with ${\cal P}$ denoting the principal value. Similarly,
\begin{eqnarray}
{\Gamma_{lmnk}^{(-)}} &=& {1 \over {8\hbar}}\left[ \Lambda^{1} J_{1}(\omega_{lm}) + \Lambda^{2}J_{2}(\omega_{lm}) \right]\left[ \coth({{\beta \hbar \omega_{lm}}\over 2}) + 1 \right] \nonumber \\ 
&+&  {i\over 4\pi\hbar}\left[ \Lambda^{2}M^{(+)}(\omega_{lm}, 2) + \Lambda^{1}M^{(+)}(\omega_{lm}, 1)\right].
\end{eqnarray}
Notice that the real parts of $R$ contribute to the decoherence and dissipation. For the purpose of studying the evolution of entanglement we will ignore the renormalization of the energy levels due to the imaginary part of the rates. This approximation has been verified to hold for a wide parameter regime for the Ising case \cite{Stor}, and for weak coupling for single qubits coupled to Ohmic baths \cite{Leggett}. 

\section{Entanglement Dynamics}

\subsection{J=0}

Consider a system of two qubits that is prepared in an entangled state, e.g., the state with eigenvalue ${1\over 8}J$, which in the absence of $\Delta$ is the $S_z = 0$ triplet state.  We will throughout this section use the terms ``singlet'' and ``triplet'' to refer to the states that are adiabatically connected to singlet and ($S_z=0$) triplet in the absence of $\Delta$.  Once initialized we allow the two qubits to evolve independently in the presence of dissipation, i.e., with $J=0$ and
nonzero $\alpha$.

To leading order in the tunneling matrix element, we can write down the equations of motion for the density matrix. Given the initial condition, the only non zero elements are governed by the following equations,
\begin{eqnarray}
\dot{\rho_{13}} &=& ({ie \over 2} - \alpha T) \rho_{13} + {\alpha n \over 4} (\rho_{22} + \rho_{33}) \nonumber \\
\dot{\rho_{22}} &=& -\alpha T\rho_{22} + \alpha T \rho_{33} - {\alpha n \over 4} (\rho_{34} + \rho_{43}) \nonumber \\
\dot{\rho_{31}} &=& -({ie \over 2} + \alpha T) \rho_{13} + {\alpha n \over 4} (\rho_{22} + \rho_{33}) \nonumber \\
\dot{\rho_{33}} &=& \alpha T\rho_{22} - \alpha T \rho_{33} 
\end{eqnarray}
The density matrix in the standard basis is 
\begin{equation}
\rho =  \left(\matrix{0&0&0&0\cr
			    0&{1\over 2}&{1\over 2}e^{-2\alpha K_{B}T t/\hbar}&0\cr
			    0&{1\over 2}e^{-2\alpha K_{B}T t/\hbar}&{1\over 2}&0\cr
			    0&0&0&0\cr}\right)
\end{equation}
where terms up to order $n/e$ have been retained. One immediately realizes that the time evolution of the density matrix is controlled by the decoherence rate of the individual qubits, $\Gamma_{\phi} = \alpha {K_{B}T \over {\hbar}}$. The dissipation is not present in the diagonal elements because they are not present at this order of approximation. The concurrence in this case is $C = e^{-2\Gamma_{\phi}t}$ and the entanglement is given by,
\begin{eqnarray}
E&=& - {{1-\sqrt{1-e^{(-4\Gamma_{\phi}t)}}}\over 2} \log_{2}({{1-\sqrt{1-e^{(-4\Gamma_{\phi}t)}}}\over 2}) \nonumber \\  &-&{{1+\sqrt{1-e^{(-4\Gamma_{\phi}t)}}}\over 2} \log_{2} ({{1+\sqrt{1-e^{(-4\Gamma_{\phi}t)}}}\over 2}).
\end{eqnarray}

\noindent The entanglement decays at a rate that is $2/\log (2)$ times faster than the decoherence rate of individual qubits as can be obtained by differentiating Eqn. $18$. The origin of this speed up can be traced back to the fact that the golden rule rates of the individual qubits are summed in the evolution equation for the density matrix and the factor of $\log(2)$ follows from the definition of entanglement. The fidelity of the state decays exponentially at a rate $2\Gamma_{\phi}$ and the final state is a incoherent mixture of $\left| \uparrow\downarrow \right>$ and  $\left| \downarrow\uparrow \right>$. Similar results are obtained if the initial state was the equivalent of the singlet or the Bell state $\left| \uparrow \uparrow\right> + \left| \downarrow\downarrow\right>$. Since we have dropped terms of higher order in $n/e$, this is not the complete story. For sufficiently long times the system reduces to a product state due to the presence of the bias $e$.

\subsection{Finite Coupling}

While the definition of entanglement can also be used for the case where the qubits remained coupled, the interpretation is less clear, as the two individual degrees of  freedom are never truly isolated from each other. Here we rely on the observation that the exponential speed up in quantum algorithms is achieved due to the inability to reduce an entangled state to a product state. This notion can be applied even to the case where the qubits remain coupled. Here too we proceed by initializing in the same eigenstate as in the previous section. As seen in Fig. $1$, for decreasing coupling the entanglement decays at a slower rate, but for larger coupling strength, the entanglement is completely lost and an incoherent state emerges at finite time. As the state evolves further, the entanglement becomes nonzero again. 

To get a better understanding of the dynamics, we consider the fidelity of the initial state (Fig. $2$). The fidelity characterizes the distance between two quantum states. It is defined as $ F \equiv$tr$\sqrt{\rho (0)^{1/2}\rho (t)\rho (0)^{1/2}}$. The entangled state evolves in a manner such that it becomes a mixture of the singlet and triplet states (note that the definition of singlet and triplet is only exact for $\Delta = 0$). Ultimately the dissipation leads to the formation of product states with zero entanglement but, for low enough temperatures, the admixture with the singlet state is nevertheless observed. For small coupling the state does not change greatly, as is reflected in the very slow decay of fidelity and entanglement. For larger values of $J$ there is a drop in fidelity but it is rather featureless, as one might expect. Entanglement on the other hand displays a far more complicated behavior. 

The behavior of entanglement can easily be understood if one looks at the effect of dissipation on the quantum state. For the given Hamiltonian, the lowest energy state of the two qubits is indeed a product state. For short times, compared to the dissipation of the diagonal elements of the density matrix elements, the ``triplet state'' has a lower energy compared to the ``singlet'' state. Thus the environment drives the system towards the singlet state leading to relaxation  (i.e. loss of off diagonal elements in the density matrix). The fidelity of the state tends to $1/2$, which implies that the final state will be found to be in the initial state in half the measurements performed. Since the state evolves from one maximally entangled state to another, via an intermediate incoherent state, the entanglement reduces to zero before starting to increase. Note that for sufficiently long times,
$O(et_{0}/n)$, the system will indeed lose all its entanglement content.

\begin{figure}
\vspace{0.2in}
\includegraphics[width=3.0in,angle=-90]{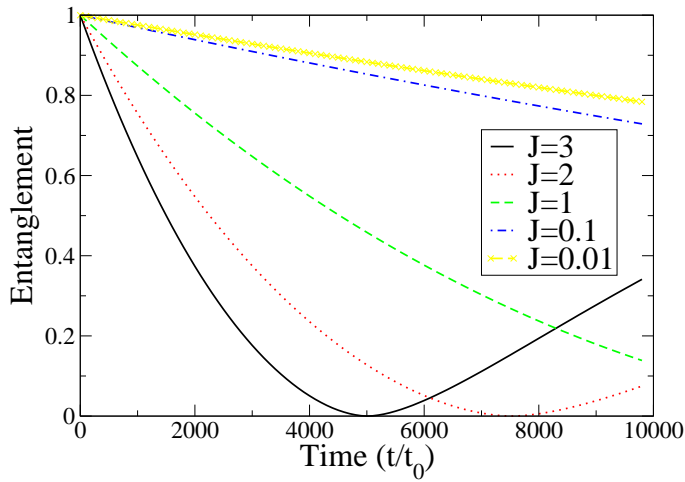}
\caption{Entanglement as a function of time starting from the triplet-like state. Time is measured in units of $t_{o}=h/e$. The parameters used here are $e = 1, n = 0.2, T = 0.1$ and $\alpha = 0.0001$.}
\label{entag}
\end{figure}

\begin{figure}\label{fidelity}
\vspace{0.2in}
\includegraphics[width=3.0in,angle=-90]{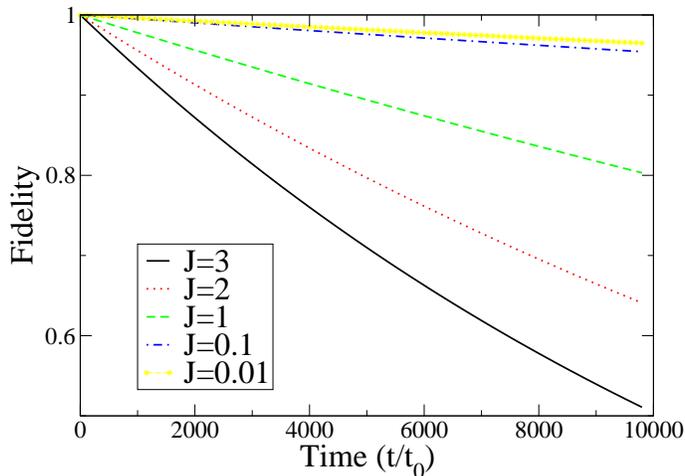}
\caption{Fidelity as a function of time for the same parameters as in Fig. \ref{entag}}
\end{figure}

To further illustrate the dynamics for long times, and the enhancement of dissipation with temperature, we now look at the time evolution of entanglement for $J = 2$. In Fig. $3$, for low temperatures, the previous dependence is reproduced but for higher temperatures we notice that the long time decay increases rapidly while the initial loss of entanglement is hardly affected. Thus the intermediate zero entanglement state is a reflection of the mixing of singlet and triplet state, whose rate is controlled by the coupling $J$, while the long time relaxation into product state is controlled by the temperature of operation.

\begin{figure}
\vspace{0.2in}
\includegraphics[width=3.0in,angle=-90]{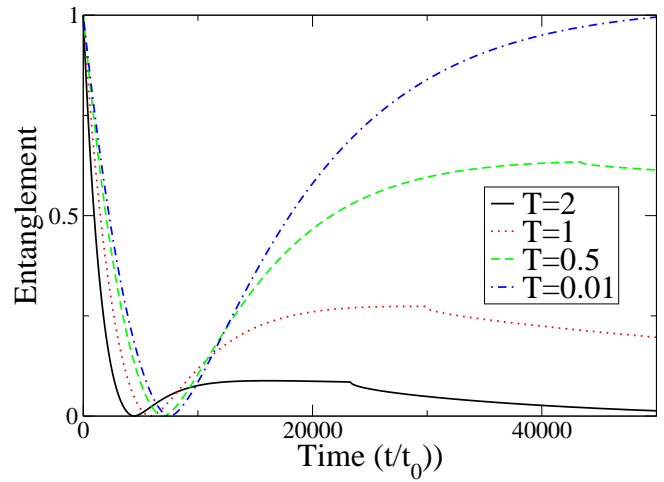}
\caption{Entanglement as a function of time. Time is measured in units of $t_{o}=h/e$. The parameters used here are $e = 1, n = 0.2, J = 2$ and $\alpha = 0.0001$.}
\label{entag1}
\end{figure}

In all cases studied here, we have considered an initial state which is is adiabatically connected to the $S_{z} = 0 $ triplet state for $\Delta = 0$. One can study the effect of dissipation on the ``singlet state'' as well. Here the dynamics is rather trivial in that one only notices a decay to a product state. This is due to the fact that at finite $J$, there is no mixing with the triplet state as was observed before, since it has a higher energy. To verify that this physical picture is correct, we have checked that the roles of the states get switched when the sign of $J$ is changed.

\subsection{Ising Coupling}

While interesting phenomena involving nonmonotonic behavior of entanglement is seen for the case where the coupling is isotropic, for an Ising-like coupling no such behavior is obtained (Fig. $4$).  We comment briefly on this case for comparison with the thorough study of gate fidelity under various operations in~\cite{Stor}.  One might have speculated that the dissipation would lead to a quantum state with a lower energy, and hence even for the Ising coupling the nonmonotonic behavior persists, but this turns out not to be the case. The decay of entanglement to zero in Fig. $1$ reflects the formation of the state $(\left|\uparrow\downarrow\right> \left<\uparrow\downarrow\right| + \left|\downarrow\uparrow\right> \left<\downarrow\uparrow\right|)/2$. The density matrix for this state commutes with the Ising coupling term in the Hamiltonian but does not do so for the $SU(2)$ symmetric coupling. Thus, in the latter case, the state continues to evolve into an entangled state. While this argument is strictly valid at zero bias, $e = 0$, the computation performed here shows that it holds approximately even for finite $e$. The dynamics shown in figure 4 is representative of all cases we have studied for difference values of the coupling $K$, both positive and negative. This is also consistent with the observation that the singlet is a protected subspace for the Ising system, in that no mixing with the triplet is observed \cite{Stor, Lidar}

\begin{figure}
\vspace{0.2in}
\includegraphics[width=3.0in,angle=-90]{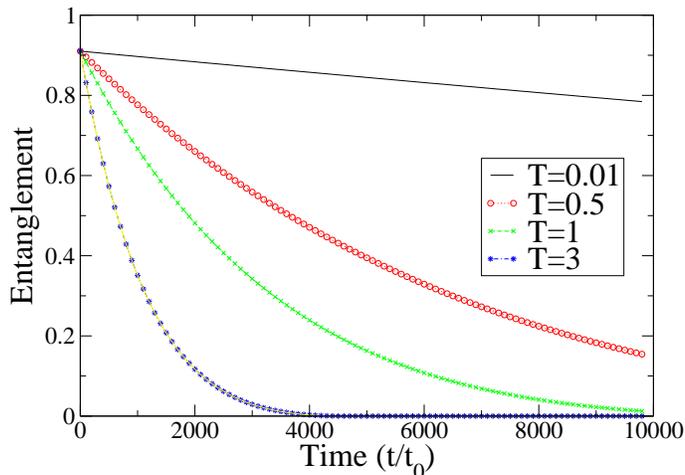}
\caption{Entanglement as a function of time for Ising coupling, starting from the entangled triplet state. Time is measured in units of $t_{o}=h/e$. The parameters used here are $ e = 1, n = 0.2, K = 1$ and $\alpha = 0.0001$.}
\label{entag2}
\end{figure}

\section{Conclusions}

In this paper we have studied the time evolution of entanglement for two-qubit systems in the presence of a dissipative environment. In the absence of any coupling between the two qubits, entanglement of the initial state is found to decay in one limit at a rate that is about three times as fast as the decoherence rates of individual qubits, and numerical results suggest that this behavior is more general. In the presence of coupling the entanglement decay rate changes dramatically and depends on the nature of the coupling itself. For the case of an Ising like coupling between the two qubits, the loss of fidelity is faithfully reproduced as a loss in entanglement. The same is not found to be true when the coupling between the two qubits is isotropic. For small coupling strengths, both entanglement and fidelity are lost rather slowly, but on increasing the the coupling not only does entanglement decay at a much faster rate, it also shows nonmonotonic behavior. In particular the entanglement is found to go to zero at intermediate times, before recovering on further time evolution. In agreement with the intuition that at higher temperatures and longer times, the system should evolve to an incoherent state, entanglement is found to vanish in this limit. 

This work is a first step in understanding the physics of coupled qubit systems in terms of their entanglement content. While the notion of entanglement of two qubits has been extensively studied and an analytic measure exists, no such understanding for multiqubit systems exists. Such an understanding is vital for implementation of efficient quantum algorithms.  It is an important task for future investigation not only to quantify entanglement for more complex systems, but also to understand its evolution in the presence of a dissipative environment.

The authors acknowledge helpful conversations with B. Plourde and T. Robertson, and support from DOE LDRD-366434 (V. A.), NSF DMR-0238760, and the Hellman Faculty Fund (J. E. M.).

\bibliographystyle{unsrt}
\bibliography{paper}

\end{document}